\documentclass[a4paper,aps,preprintnumbers,twocolumn,amsmath,amssymb,prl,showpacs]{revtex4}
\usepackage{bbm}
\usepackage{mathrsfs}
\usepackage{amsmath}
\usepackage{nicefrac}
\usepackage[dvips]{graphicx}
\usepackage{epsfig}

\newcommand {\ud}  {\mathrm{d}}
\renewcommand{\(}{\left(}        \renewcommand{\)}{\right)}
\renewcommand{\*}{\cdot}
\renewcommand{\/}[2]{\frac{#1}{#2}}
\newcommand{\bra}{\langle}       \newcommand{\ket}{\rangle}
\renewcommand{\vec}[1]{\boldsymbol{#1}}
\newcommand{\mc}{\mathcal}

\newcommand{\Tr}{\operatorname{Tr}}
\newcommand{\Vol}{\operatorname{Vol}}
\newcommand{\supp}{\operatorname{supp}}
\newcommand{\transp}{\intercal}
\newcommand{\pdag}{{\phantom\dag}}

\newcommand{\sys}{\Omega}
\newcommand{\vareps}{\varepsilon}

\begin{document}
\title{Dephasing and the steady state in quantum many-particle systems}

\author{T. Barthel and U. Schollw\"ock}
\affiliation{Institute for Theoretical Physics C, RWTH Aachen, D-52056 Aachen, Germany}

\begin{abstract}
We discuss relaxation in bosonic and fermionic many-particle systems. For integrable systems, the time evolution can cause a dephasing effect, leading for finite subsystems to certain steady states. We give an explicit derivation of those steady subsystem states and devise sufficient prerequisites for the dephasing to take place. We also find simple scenarios, in which dephasing is ineffective and discuss the dependence on dimensionality and criticality. It follows further that, after a quench of system parameters, bipartite entanglement entropy will become extensive. This provides a way of creating strong entanglement in a controlled fashion.
\end{abstract}

\date{March 12, 2008}

\maketitle

In equilibrium statistical physics we usually work with canonical ensembles, characterized by density matrices $\hat{\varrho} \propto e^{-\sum_k \alpha_k \hat O_k}$ that can be derived by maximizing the entropy under the constraint of fixed expectation values $\bra\hat O_k\ket$ for a few (macroscopic) observables like energy or particle number \cite{Jaynes1957-106}. 
However, it is in general unclear whether a given system in a certain initial state will evolve to any steady state at all and if so, whether that state is indeed one of the canonical ensembles. Recent experiments, especially with ultracold gases, have revived the interest in this important topic of nonequilibrium physics. In such experiments with integrable systems, absence of thermalization was observed (e.g.\ in \cite{Kinoshita2006-440}). We will deduce analytically under what circumstances integrable systems relax to non-canonical steady states. 

It was conjectured in \cite{Rigol2007-98} that the time evolution in an integrable system with conserved observables $\hat I_k$ should lead to the corresponding maximum entropy ensemble \cite{Jaynes1957-106}
\begin{equation}\textstyle
\label{eq:RigolConjecture}
\tilde \varrho_d=\/{1}{Z}e^{-\sum_k\alpha_k\hat I_k}\,, 
\end{equation} 
where the $\alpha_k$ are determined by the initial state. So far, the conjecture was discussed by analyzing specific observables for specific systems  \cite{Rigol2007-98,Rigol2006-74,Cazalilla2006-97,Eckstein2007}. In \cite{Gangardt2007} it was found that certain observables do not relax to the ones predicted by the ensemble $\tilde \varrho_d$ as conjectured in \cite{Rigol2007-98}. So here the focus will be not on observables but states themselves.

We point out that it is essential, for the conjecture to hold, to restrict oneself to measurements in a finite subsystem, i.e.\ \eqref{eq:RigolConjecture} should not be interpreted as the steady state of the full system (see a first discussion in \cite{Cramer2007}). A general proof is given, showing explicitly how state relaxation in such subsystems can occur due to dephasing. We devise the necessary prerequisites, discuss the relaxation speed, and give simple examples and counter examples for dephasing. Our results allow for a simple interpretation of the discrepancies between \cite{Rigol2007-98} and \cite{Gangardt2007}.
On general grounds it follows further that through dephasing the bipartite entanglement entropy in a pure state will become extensive. This may be of interest for quantum computation applications where entanglement is needed as a resource. We comment on implications for the ability to simulate  such systems on classical computers. We will first present the case of free systems, then point out how this generalizes to Bethe Ansatz integrable models and close with a short discussion of nonintegrable systems.

\emph{Dephasing for quadratic Hamiltonians.--}
First, we address the situation of quadratic Hamiltonians (for $t>0$),
\begin{equation}\textstyle 
\label{eq:Hquadratic}
        \hat H = \sum_{ij} [ a_i^\dag V_{ij}a_j^\pdag + \/{1}{2} 
         (a_i^\dag W_{ij} a_j^\dag + h.c. )]+const.\,,
\end{equation} 
where $a_i$ are bosonic or fermionic ($\zeta:=\pm 1$) ladder operators, $[a_i^\pdag,a_j^\dag]_{-\zeta}=\delta_{ij}$, and $\vec{a}^\transp=(a_1,a_2,\dotsc)$. This covers also lattice regularized free field theories. The Hamiltonian is diagonalized by a linear canonical transformation
\begin{equation}\textstyle
\binom{\vec{a}}{\vec{a}^\dag}=U\binom{\vec{\eta}}{\vec{\eta}^\dag},\quad
\hat H=(\vec{\eta}^\dag)^\transp\vareps\vec{\eta}^\pdag=\sum_k\vareps_k \eta_k^\dag\eta_k^\pdag\,,
\end{equation}
where $\vareps$ is the diagonal matrix of one-particle energies.
For quadratic systems, the groundstate, thermal or evolved states are all Gaussian. Assuming such a Gaussian initial state $\hat \varrho_{t=0}$, due to the Wick theorem, the state is always fully characterized by the one-particle Green's functions $\mc{G}$ (superscripts $a,\eta$ indicate the chosen basis)
\begin{equation}\textstyle
\label{eq:GreenFct}
\hat \varrho=\hat \varrho(\mc{G})\,,\quad
\mc{G}^a\equiv\bra \binom{\vec{a}}{\vec{a}^\dag}\*\binom{\vec{a}^\dag}{\vec{a}}^\transp\ket_{\hat \varrho}
=U\mc{G}^\eta U^\dag\,.
\end{equation}
 The initial state $\hat \varrho_{0}$ might e.g.\ be the groundstate of a different quadratic Hamiltonian (``quench'' of system parameters at $t=0$). In the following, we will consider bipartitions of the full system into a subsystem $\sys$ and its environment $\sys^\bot$, call $\Theta_\sys$ the projection onto $\sys$, and the volume $\mc V:= \Vol\Omega+\Vol\Omega^{\bot}$. In the situation where $\sys^\bot$ is approaching the thermodynamic limit, the quantum number labels $k$ become a set of $d$ continuous labels and one discrete label $k\to(\vec{k},s)$ (e.g.\ a $d$-dimensional momentum vector and a band index, $ \vareps_k=\vareps_{\vec{k}s}$ denoting the dispersion relation) with  $\vec{k}\in \Gamma$, $\Vol\Gamma$ finite, and the density of states $\rho_s\propto 1/{\mc V}$. 
Again, due to the Wick theorem, reduced density matrices $\Tr_{\sys^\bot}\hat \varrho$ of quadratic systems are functions of the one-particle subsystem density matrices $G$, $\Tr_{\sys^\bot}\hat \varrho=\hat\varrho(G)$, where $G$ is defined by $G^a= \Theta_\sys\mc{G}^a\Theta_\sys$. 

\emph{Theorem.}
Let the Green's function $\mc{G}_d$ be defined by 
\begin{equation}
[\mc{G}_d^\eta]_{kk'}:=\delta_{kk'}[\mc{G}_{t=0}^\eta]_{kk}\,,
\end{equation}
 i.e., in the eigenbasis representation, $\mc{G}_d$ is the diagonal part of $\mc{G}_{t=0}$. 
Under preconditions \emph{(a-c)}, stated after the proof, the steady $t\to\infty$ state of $\sys$ is then given by
\begin{equation}
\label{eq:TheoremQuadratic}
\lim_{t\to\infty}\Tr_{\sys^\bot} \hat\varrho_t
= \Tr_{\sys^\bot}\hat{\varrho}_d
\quad\text{with}\quad
\Tr_{\sys^\bot}\hat{\varrho}_d= \hat \varrho({G}_d)\,,
\end{equation}
i.e.\ subsystem states relax to the reduced density matrices of $\hat \varrho_d=\tilde \varrho_d$, the maximum entropy ensemble \eqref{eq:RigolConjecture} with
\begin{equation}\textstyle
\label{eq:alpha}
 \hat{I}_k=\eta_k^\dag\eta_k^\pdag\quad\text{and}\quad \alpha_k = \ln \(\zeta/(1-[\mc{G}_d^\eta]_{kk}^{-1})\).
\end{equation}

\begin{figure}[b]
\epsfig{file= 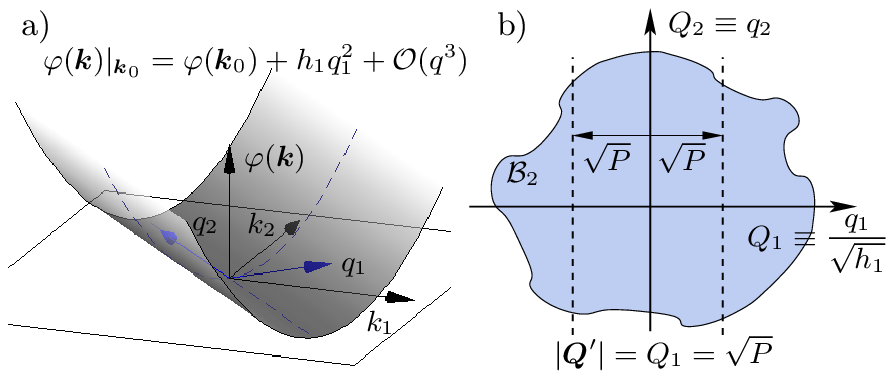, angle=0,width=1\linewidth}
\caption{Sketch of coordinate transformation and integration path in \eqref{eq:GreenFourierCritical} for a non-isolated zero of $\vec{g}$  with $d=2$, $n=1$.} \label{fig:dephasing_critPoint.FIGG}
\end{figure}

\emph{Proof.}
To compare the two reduced density matrices in proposition \eqref{eq:TheoremQuadratic}, we compare at first the corresponding subsystem Green's functions $\lim_{t\to\infty}G_t$, $G_d$. They are 
\begin{gather}
\label{eq:GdiagSubsys}
G_d^a=\Theta_\sys(U\mc{G}_d^\eta U^\dag)\Theta_\sys\,, \quad \text{and}\\
\textstyle
G_t^a = \Theta_\sys\bra\hat u_t^\dag \binom{\vec{a}}{\vec{a}^\dag}\binom{\vec{a}^\dag}{\vec{a}}^\transp \hat u_t \ket_{\hat\varrho_0} \Theta_\sys 
= \Theta_\sys U u_t^\eta\, \mc{G}_0^\eta\, (u_t^\eta)^\dag\, U^\dag  \Theta_\sys,
\notag
\end{gather}
where $\hat u_t=e^{\hat H t/i\hbar}$, $u_t^\eta=e^{\mathcal{E} t/i\hbar}$, 
$\mathcal{E}:=(\begin{smallmatrix}\vareps&\\&-\vareps\end{smallmatrix})$, and thus
\begin{gather}
\textstyle
\label{eq:GtSubsys}\textstyle
G_t^a = \/{1}{\mc V}\sum_{kk'} e^{(\mathcal{E}_k-\mathcal{E}_{k'})t/i\hbar} f_{kk'}\,,
\quad\text{with}\\
 \label{eq:integrand}\textstyle
 f_{kk'}:=\mc V \*\Theta_\sys \vec{U}_k \,[\mc{G}_0^\eta]_{kk'}\, \vec{U}^{\dag}_{k'} \Theta_\sys,\quad
 [\vec{U}_k]_i:=U_{ik} \,.
\end{gather}
Here,  $f_{kk'}$ is a matrix valued function of $k$ and $k'$, whose matrix indices label points in  $\Omega$.
Comparing $G_t^a$ 
to $G_d^a$, 
we see that \eqref{eq:TheoremQuadratic} is true, if the nondiagonal contributions $k\neq k'$ to $G_t^a$ 
 vanish for $t\to\infty$. The summation over $k$, for a fixed $\Delta k = k-k'\neq 0$, corresponds for increasing $t$ to a Fourier transform with respect to ever higher frequencies which may vanish due to phase averaging (hence we call the effect ``dephasing''): To see this, let us rewrite $G_t^a$, \eqref{eq:GtSubsys}, with $(k,k')\to({\vec{k},s,\vec{k}'=\vec{k}+\Delta\vec{k},s'})$ as 
\footnote{One needs to exclude cases where single modes give, in the thermodynamic limit, finite contributions to the sum.}
\begin{equation}
\label{eq:GreenFourier}
G_t^a \to \sum_{ss'}\/{1}{\mc V\rho_s} \sum_{\Delta\vec{k}} 
\int\ud^d{k}\, e^{-i\varphi^{ss'}_{\Delta\vec{k}}(\vec{k})t} f_{\Delta\vec{k}}^{ss'}(\vec{k})\,,
\end{equation}
with phase function $\varphi$ and group velocity difference $\vec g$,
\begin{equation}
 \label{eq:phasefct}\textstyle
 \varphi^{ss'}_{\Delta\vec{k}}(\vec{k}) :=
  \/{\mathcal{E}_{\vec{k}s}-\mathcal{E}_{(\vec{k}+\Delta\vec{k})s'}}{\hbar}\,,\quad
 \vec{g}^{ss'}_{\Delta\vec{k}}(\vec{k}):=
 \partial_{\vec k} \varphi^{ss'}_{\Delta\vec{k}}(\vec{k}).
\end{equation}
In the following, we omit indices $s$, $s'$, and $\Delta\vec{k}$ and consider always nondiagonal contributions to \eqref{eq:GreenFourier}, i.e.\ $s\neq s'$ or $\Delta\vec{k}\neq\vec{0}$.
If $\varphi\in \mc C^1$, $\vec{g}$ finite on $\supp(f)\subset \Gamma$, and if the matrix elements of $f$ are $\mc L^1$ integrable 
\footnote{$f\in \mc L^1$ means that $\int \ud^d k |[f(\vec{k})]_{ij}|$ exists $\forall_{ij}$.},
the integral of \eqref{eq:GreenFourier} vanishes for $t\to\infty$  (Riemann-Lebesgue Lemma). This is due to the fast oscillation of the Fourier kernel $e^{ -i\vec{g}\*(\vec{k}-\vec{k}_0)\,t}$ in the vicinity of every point $\vec{k}_0$. Further, if $f$ is bounded in a vicinity $\mc B$ of $\vec{k}_0$, the contribution of $\mc B$ to the integral is of $\mc O(\Vol\mc B/t)$.
But what happens at points $\vec{k}_0$ where $\vec{g}$ vanishes? If such a zero of $\vec{g}$, is isolated (i.e.\ if the Hesse matrix of $\varphi$ at $\vec{k}_0$ is invertible) and if $f$ is smooth in a vicinity of $\vec{k}_0$, its contribution to \eqref{eq:GreenFourier} is still vanishing (follows from the Morse lemma and a Gaussian integral).
We can also treat the case where the Hesse matrix $\mathfrak{H}_\varphi:=\/{\partial}{\partial\vec{k}}(\/{\partial}{\partial\vec{k}})^\transp\varphi|_{\vec{k}_0} =:W^\transp h W$ has only n ($1\leq n\leq d$) nonzero eigenvalues $\{h_i\}_{1\leq i \leq n}$. With the transformations $\mc M_1:\vec{k}\to (\vec{q}',\vec{q}''):=\vec{q}= W\*(\vec{k}-\vec{k}_0)$ ($\vec{q}'\in \mathbb R^n$, $\vec{q}''\in \mathbb R^{d-n}$) and $\mc M_2:\vec{q}\to (\vec{Q}',\vec{Q}'')=(h^{-1/2}\vec{q}',\vec{q}'')$, the contribution from a vicinity $\mc B$ of $\vec{k}_0$ becomes (see Fig.~\ref{fig:dephasing_critPoint.FIGG})
\begin{multline}
\label{eq:GreenFourierCritical}\textstyle
\int_{\mc{B}}\ud^d{k} e^{-i\varphi(\vec{k})t} f(\vec{k})
=\int_{\mc{B}_2}\ud^d{Q}\, e^{-it\sum Q_i'^2+\mc O(Q^3)} f_2(\vec{Q})\\\textstyle
\sim \int \ud P \, e^{-it P} F(P)\,,
\end{multline}
where $\mc B_2=\mc M_2(\mc M_1(\mc B))$, $f_2=f\circ\mc M_1^{-1}\circ\mc M_2^{-1}$, and 
$F(P) := \/{1}{2 \sqrt{P}}\int_{\mc{B}_2,|\vec{Q}'|=\sqrt{P}} \ud^{d-1}{Q}\, f_2(\vec{Q})$.
Integral \eqref{eq:GreenFourierCritical} has again the form of a Fourier transform and vanishes for $t\to\infty$, if $F\in\mc L^1( \mc B_2)$. If $F$ is bounded in the vicinity of $P=0$, the integral vanishes as $\mc O(\Vol\mc B/t)$. Those conditions on $F$ have to be interpreted as stricter conditions on $f$ when approaching zeros of $\vec{g}$, appropriate to still guarantee the dephasing. Zeros of $\vec{g}$ with multiplicity $\ell-1>0$ ($\mathfrak{H}_\varphi=0$) can be treated in a general fashion only for 1d by substituting $Q=q^\ell$. In higher dimensions they are a lot more complicated; see e.g.\ \cite{Varchenko1976} for 2d.

In the cases discussed so far, we have assumed that $\varphi$ is continuously differentiable ($\varphi\in\mc C^1$). For the sake of brevity we will consider now only the scenario $\varphi(\vec{k})=\varphi_0+|\vec{k}|^\ell+\mc O(k^{\ell+1})$. This is for $\ell=1$ nondifferentiable. The cases $\ell=1$ and $\ell=2$ cover very typical examples (for example magnons, phonons or free particles). The contribution of a (small) sphere of radius $K$ around $\vec{k}_0=\vec{0}$ to the integral in \eqref{eq:GreenFourier} is with the substitution $P=k^\ell$
\begin{equation}
\approx \int_0^{K^\ell}\hspace{-1em} dP\, e^{i(\varphi_0+P)t}\tilde F(P),\quad
\tilde F(P):=\/{P^{\/{1-\ell}{\ell}}}{\ell }\int_{|\vec{k}|=P^{\/{1}{\ell}}}\hspace{-2.2em} \ud^{d-1} k\,  f(\vec{k}).
\end{equation}
This contribution vanishes for $t\to\infty$ if $\tilde F \in\mc L^1_{loc}$ and if $\tilde F$ is bounded, the contribution is of $\mc O(K^d/t)$.

For a few general, physically relevant scenarios, we have established sufficient conditions under which (every matrix element of) the integral in \eqref{eq:GreenFourier} goes to zero as $t\to\infty$. As $\Vol \Gamma$ is finite and as we have a finite number of bands $s$, those conditions guarantee hence that all nondiagonal contributions to $G_t^a$ vanish.
Due to Wick's theorem, expectation values of arbitrary observables $\hat O$ on $\sys$ are given by polynomials in $[G_t^a]_{ij}$. Nondiagonal contributions to $\bra\hat O\ket$ will consequently also vanish if we restrict to finite subsystem sizes. 
From the convergence $G_t^a\to G_d^a$ follows then the proposition \eqref{eq:TheoremQuadratic}. Finally, \eqref{eq:alpha} follows from $\bra\eta_k^\dag\eta_k^\pdag\ket_{\tilde\varrho_d}=1/(e^{\alpha_k} - \zeta)$ and $\zeta\bra\eta_k^\dag\eta_k^\pdag\ket_{\hat\varrho_d}=[\mc{G}_d^\eta]_{kk}-1$.
\hfill$\Box$

\emph{Preconditions.}
During the proof, we collected the following prerequisites, sufficient to guarantee convergence to the steady state \eqref{eq:TheoremQuadratic}: 
\emph{(a)} $\Vol\sys$ is finite and $\mc V\to\infty$.
\emph{(b)} The parameterization $k\to(\vec{k}\in\Gamma,s)$ of the quantum numbers is possible with a finite $\Vol \Gamma$ and a finite number of bands $s$.
\emph{(c.1)} $\varphi\in \mc C^1$, $\vec{g}$ finite on $\supp f\subset\Gamma$, and $f\in \mc L^1$. If this is not given for the vicinity of a point $\vec{k}_0$, we require for such a point
\emph{(c.2)} if $\vec{g}(\vec{k}_0)=\vec{0}$, then the Hesse matrix $\mathfrak{H}_\varphi|_{\vec{k}_0}$ exists and is nonzero, and $F\in \mc L^1$, or
\emph{(c.3)} if, at $\vec{k}_0=\vec{0}$, $\varphi(\vec{k})=\varphi_0+|\vec{k}|^\ell+\mc O(k^{\ell+1})$ then $\tilde F\in \mc L^1$. Of course, \emph{(c.3)} can be generalized to the case $|\vec{k}|^\ell\to |A\*(\vec{k}-\vec{k}_0)|^\ell$ with some nonzero matrix $A$.

With those conditions, all nondiagonal contributions to the subsystem Green's function matrix $G^a_t$ vanish for $t\to\infty$ and dephasing to the steady state ensemble is effective, \eqref{eq:TheoremQuadratic}. If $f$, $F$, and $\tilde F$ are, in the corresponding situations, bounded instead of only $\mc L^1$ integrable, nondiagonal contributions to $G^a_t$ decay (more quickly) as $\mc O(1/t)$. Below, we give illustrative examples. Among those are simple scenarios where some of the prerequisites are violated and dephasing does in fact not occur.

\emph{Examples and counter-examples for dephasing.--}
At the end of the proof, a reason for requiring \emph{(a)} was given. A simple counter-example consists in violating \emph{(a)} with $\sys^\bot=\emptyset$ and measuring $\zeta \bra\eta_{k'}^\dag\eta_{k}^\pdag\ket_{\hat \varrho_t } = e^{-i\varphi_{kk'}t} [\mc G_0^\eta]_{kk'} -\delta_{kk'}$, i.e.\ measurements in infinite subsystems, can  reveal the phases and nondiagonal contributions (``rephasing''). 

\begin{figure}[bt]
\epsfig{file= 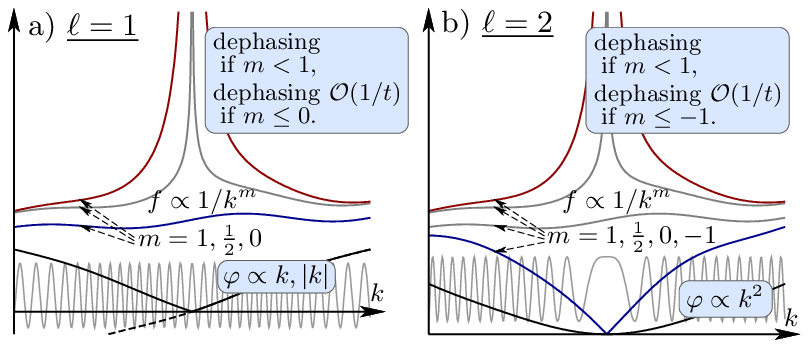, angle=0,width=1\linewidth}
\caption{Phase function $\varphi$, $\sin(\varphi(k) t)$, and the nondiagonal
 contribution $f$ to $G^a_t$ in the paradigmatic case \eqref{eq:example_paradigm}, $d=1$.}
\label{fig:dephasingCases_isotrope.FIGGG}
\end{figure}

If $\vec{g}$ has zeros or if $f$ has divergences, dephasing properties are dominated by the vicinities of such points. Thus, we illustrate \emph{(c)} by considering the paradigmatic scenario $\varphi(\vec{k})\sim \varphi_0+|k|^\ell$, $f(\vec{k})\sim 1/k^m$ near $\vec{k}=\vec{0}$ (for some fixed $s,s',\Delta\vec{k}$ and $i,j\in\Omega$). The integral in \eqref{eq:GreenFourier} is then
\begin{equation}
\label{eq:example_paradigm}\textstyle
e^{i\varphi_0 t}\int\ud^d k\/{1}{|\vec{k}|^m} e^{i|\vec{k}|^\ell t}
\sim \int\ud q \/{1}{q^\chi} e^{i q t},
\quad \chi=\/{m+\ell-d}{\ell}.
\end{equation}
Hence this (nondiagonal) contribution to $G^a_t$, for $t\to\infty$,
does not vanish if $\chi\geq 1$, 
vanishes as $1/{t^{1-\chi}}$ if $0<\chi<1$, 
and (at least) as $1/t$ if $\chi<0$; 
see Fig.~\ref{fig:dephasingCases_isotrope.FIGGG}.
In this scenario, both \emph{(c.2)} and \emph{(c.3)} apply with $F,\tilde F\propto P^{-\chi}$ and are not only sufficient but also necessary.

\begin{figure}[bt]
\epsfig{file= 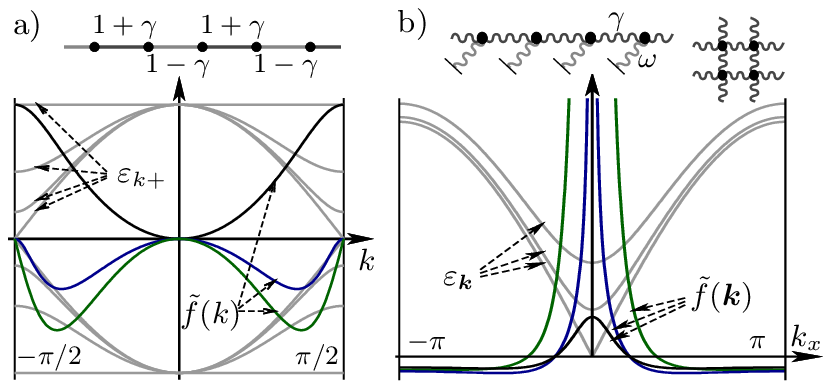, angle=0,width=1\linewidth}
\caption{a) Dimerized fermionic 1d tight-binding model \eqref{eq:HamThightBindDimer}. Dispersion relation $\vareps_{k\pm}=\mp\varphi(k)/2$ for  $\gamma=1,\/{1}{2},\/{1}{5},0$ and the nondiagonal contribution $\tilde f(k)$ to $\bra c_1 c_0^\dag\ket_t$ according to Eq.~\eqref{eq:dephasingThightBindDimer} for quenches $\gamma=\/{1}{2}\to 0,\,\/{1}{5}\to \/{1}{2},\,0\to \/{1}{2}$ (top to bottom).
b) The $d$-dimensional harmonic lattice model. Dispersion relation $\vareps_{\vec{k}}=-\varphi(\vec{k})/2$ for $\omega=\/{4}{5},\/{2}{5},0$ with $\gamma=1$, and the nondiagonal contribution $\tilde f(\vec{k})$ to $\bra b_{\vec{x}} b_{\vec{x}'}^\dag\ket_t$ according to Eq.~\eqref{eq:dephasingHarmonicLattice} for quenches $\omega=\/{4}{5}\to 0,0\to \/{4}{5},\/{2}{5}\to \/{4}{5}$ (top to bottom).} \label{fig:examples.FIGG}
\end{figure}

Our first explicit example is ($a_i\equiv c_x$, $\hbar=1$)
\begin{equation}\textstyle
\label{eq:HamThightBindDimer}
\hat H=-\sum_{x} (1+\gamma(-1)^x)[c_x^\dag c_{x+1}^\pdag+h.c.]\,,
\end{equation}
the dimerized fermionic tight-binding model,
where modes $c_k$ and $c_{k+\pi}$ are coupled and the dispersion relation is $\vareps_{k\pm}=\pm 2\sqrt{\cos^2 k+\gamma^2\sin^2 k}$, i.e.\ gapless if $\gamma=0$. The eigenmodes are labeled $\eta_{k\pm}$.
We evolve the groundstate for a certain dimerization $\gamma_0$ with a different value $\gamma\neq\gamma_0$. Skipping details of the calculation, we note that, using \eqref{eq:GreenFourier}, the nondiagonal contributions $\bra c_x^\pdag c_{x'}^\dag\ket_t^{nd}$ from $\bra\eta_{k+}^\pdag\eta_{k-}^\dag\ket$ and $\bra\eta_{k-}^\pdag\eta_{k+}^\dag\ket$ to $\bra c_x^\pdag c_{x'}^\dag\ket_t$ can be written as
\begin{equation}\textstyle
\label{eq:dephasingThightBindDimer}
\bra c_x^\pdag c_{x'}^\dag\ket_t^{nd} = \int_0^{\/{\pi}{2}}\ud k \tilde f(k)
\*\begin{cases}
\cos(\varphi(k)t), &\hspace{-0.4em}\text{odd }x-x'\\
i\sin(\varphi(k)t),&\hspace{-0.4em}\text{even }x-x'
\end{cases}
\end{equation}
Fig.~\ref{fig:examples.FIGG}a displays $\vareps_{k\pm}=\mp\varphi(k)/2$ and $\tilde f$ for $(x,x')=(1,0)$ and three different quenches with $\gamma\neq 1$. The zeros of $g$ are $k=0,\pm\/{\pi}{2}$ and have each, in the notation of the paradigmatic situation \eqref{eq:example_paradigm}, one of the characteristics $(l,m)=(2,-2),(1,0),(2,-1)$. Hence, $\chi< 0$ and dephasing of $\mc O(1/t)$ is guaranteed. The same is given for all other $(x,x')$ and has also been checked numerically. However, if we switch to $\gamma=1$, $\varphi$ is const. $=-4$ $\forall_{k}$  and no dephasing can occur -- we have uncoupled dimers.

As a second explicit example we choose the harmonic lattice model in $d$ dimensions ($a_i\equiv b_{\vec{r}}$, $\hbar=1$). Contrary to the first example, it will not dephase in all cases.
\begin{equation}\textstyle
\label{eq:HamHarmonicLattice}
\hat H=\/{1}{2}\sum_{\vec{r}} [P_{\vec{r}}^2+\omega^2 Q_{\vec{r}}^2+\gamma\sum_{i=1}^d(Q_{\vec{r}}-Q_{\vec{r}+\vec{e}_i})^2]\,.
\end{equation}
The dispersion relation is $\vareps_{\vec{k}} = \sqrt{\omega^2+4\gamma\sum_{i=1}^d \sin^2 k_i/2}$, i.e.\ gapless for $\omega=0$.
With the bosonic operators $b_{\vec{r}}:=(Q_{\vec{r}}+iP_{\vec{r}})/\sqrt{2}$, $\hat H$ is brought to the form \eqref{eq:Hquadratic} and is hence amenable to the theorem. Using \eqref{eq:GreenFourier} we arrive at
\begin{gather}\textstyle
\bra b_{\vec{r}}^\pdag b_{\vec{r}'}^\dag\ket_t^{nd} = \int_{-\pi}^{\pi}\ud^d k f(\vec{k})\cos(2\vareps_{\vec{k}}t),\quad f=e^{i\vec{k}\*(\vec{r}-\vec{r}')}\tilde f,\notag\\ \textstyle\label{eq:dephasingHarmonicLattice}
\tilde f = \/{1}{16}({1}/{\vareps}-\vareps)(\alpha^2-{1}/{\alpha^2}),\quad
\alpha_{\vec{k}}=\sqrt{{\vareps_{\vec{k}}}/{\vareps_{\vec{k}}'}},
\end{gather}
where we switch at $t=0$ the oscillator frequency $\omega'\to\omega$ and
$\vareps_{\vec{k}}'$ is the dispersion relation before that quench. The dephasing properties are dominated by the vicinity of $\vec{k}=\vec{0}$. 
If one switches between two noncritical values $\omega',\omega>0$, its characteristic in terms of \eqref{eq:example_paradigm} is $(l,m)=(2,0)$, i.e.\ $\chi=\/{2-d}{2}$ and hence dephasing of $\mc O(1/\sqrt{t})$ for $d=1$ and $\mc O(1/t)$ for $d>1$. 
For $\omega'=0$, $\omega>0$, one has $(l,m)=(2,1)$ and consequently no dephasing for $d=1$, and dephasing of $\mc O(1/\sqrt{t})$ [$\mc O(1/t)$] for $d=2$ [$d=3$].
For $\omega'>0$, $\omega=0$, one has $(l,m)=(1,2)$ and hence no dephasing for $d=1,2$, and dephasing of $\mc O(1/t)$ for $d=3$; Fig.~\ref{fig:examples.FIGG}b. This was confirmed numerically for $d=1,2$.

As a last example, consider free hard-core bosons in a 1d box. The system is prepared in the groundstate for a box of size $\tilde L$ which is switched to $L>\tilde L$ at $t=0$, \cite{Rigol2007-98}. The Jordan-Wigner transformation yields a model of free fermions. The transformation between one-particle eigenstates before and after the quench ($|q\ket$ and $|k\ket$) is 
\begin{equation}\textstyle
 V_{kq}=\bra k|q\ket
=\/{e^{i(k-q)/2}}{\sqrt{L\tilde L}} \/{\sin(\tilde L(k-q)/2)}{\sin((k-q)/2)}
=:V_{k-q}.
\end{equation}
The weight of $V_{\Delta k}$ is concentrated in the interval $|\Delta k|\lesssim 2\pi/\tilde L$. With the Fermi momentum $q_F$, the initial Green's function $[\mc G_0^{\tilde \eta}]_{qq'}=\delta_{qq'}(1-\theta(q_F-|q|))$ is diagonal in the $|q\ket$ basis and $[\mc G_0^{\eta}]_{kk'}=[V\mc G_0^{\tilde\eta}V^\dag]_{kk'}$, which appears in \eqref{eq:integrand}, is hence also concentrated in $|k-k'|\lesssim 1/\tilde L$. Thus $|\varphi_{\Delta k}(k)|$ and $|g|$ are $\lesssim 1/\tilde L$ and dephasing is ineffective. In \cite{Rigol2007-98}, the bosonic momentum distribution $\bra\hat n_k\ket$ was found to relax to the one of the corresponding steady state ensemble $\hat \varrho_d$. However, as the dephasing is ineffective, relaxation \eqref{eq:TheoremQuadratic} of subsystem density matrices does not occur. This is also visible in the observables: As derived in \cite{Gangardt2007}, correlators do not relax to the value predicted by $\hat \varrho_d$; see also non-decaying oscillations of $\bra\hat n_x\ket$ in Ref.~41 of \cite{Rigol2007-98} and Fig.~10 of \cite{Rigol2006-74}. That a particular observable, here $\bra\hat n_k\ket$, may relax anyway is a different issue. In \cite{Gangardt2007} it was shown for a slightly modified setup how relaxation of $\bra\hat n_k\ket$ occurs.

\emph{Discussion.--}
The dephasing theorem \eqref{eq:TheoremQuadratic} confirms the conjectured \eqref{eq:RigolConjecture}, clarifies its interpretation, and devises conditions for its applicability. Dephasing properties are determined in particular by points where the gradient $\vec{g}=\partial_{\vec{k}}\varphi$ of the phase function, \eqref{eq:phasefct}, vanishes or the amplitude $f$, \eqref{eq:integrand}, diverges.
Also note that the notion of integrals of motion standing in involution as used for classical systems does not carry over to quantum mechanics, \cite{Weigert1992-56}. Hence it was per se not clear what operators $\hat I_k$ were to be chosen in the maximum entropy ensemble \eqref{eq:RigolConjecture}. The theorem settles this question. 
Further, as $\alpha_k$, \eqref{eq:alpha}, becomes finite for finite $k$ regions, the subsystem (entanglement) entropies will finally be dominated by the extensive contribution ${\Vol\sys}\sum_k \log[ (1+\zeta  e^{-\alpha_k})^{-\zeta}]$ (cmp.\ to \cite{Calabrese2005} for 1d).
Hence, the required computational resources to simulate such systems on classical computers scale exponentially in the system size, preventing access to arbitrarily long times. On the other hand, this shows that quenches are a simple tool for the controlled generation of strong (extensive) entanglement.

\emph{Bethe Ansatz integrable systems.--}
In Bethe Ansatz solvable models 
\cite{Zachary1996},
the transfer matrix $\hat \tau(\lambda)$ is conserved for any value of the spectral parameter $\lambda$;
$[\hat \tau(\lambda),\hat \tau(\lambda')]=0$ and
$[\hat \tau(\lambda),\hat H]$ $\forall_{\lambda,\lambda'}$.
Initial states $\hat\varrho_0$ can be expanded in a $\hat\tau(\lambda)$-eigenbasis $|\vec{\lambda}\ket$  and, via time evolution, nondiagonal contributions will attain quickly oscillating phases
$\hat\varrho_t=\sum_{\vec{\lambda},\vec{\lambda}'} e^{(E_{\vec{\lambda}}-E_{\vec{\lambda}'})t/i\hbar} |\vec{\lambda}\ket\bra\vec{\lambda}| \hat\varrho_0 |\vec{\lambda}'\ket\bra\vec{\lambda}'|$.
It will be shown elsewhere that, as in the free case, the nondiagonal contributions to the density matrix $\Tr_{\sys^\bot}\hat \varrho_t$ of a finite subsystem $\sys$ will under appropriate preconditions decay. Then, the steady state in the thermodynamic limit will, in generalization of \eqref{eq:TheoremQuadratic}, be given by
\begin{equation*}
\label{eq:TheoremBethe}\textstyle
\lim_{t\to\infty}\Tr_{\sys^\bot} \hat\varrho_t
= \Tr_{\sys^\bot}\hat{\varrho}_d\,,\quad
\hat{\varrho}_d = \/{1}{Z}e^{-\int\ud\lambda\, \rho(\lambda) \alpha_\lambda\hat\tau(\lambda)}\,,
\end{equation*}
where $\rho$ denotes the density of quasiparticles. 

\emph{Nonintegrable systems.--}
Whether or how thermalization occurs in nonintegrable systems is in general unclear.
Intuitively, information about the initial state gets smeared out by scattering events which are, contrary to the integrable case \cite{Mussardo1992-218}, able to change the quantum numbers of the involved particles and not factorizable.
Our results are expected to carry over to nonintegrable cases, if system and initial state allow for a description by an integrable theory of quasiparticles (e.g.\ Fermi gases and Luttinger liquids) and quasiparticle lifetimes exceed time scales necessary to observe dephasing.  In such cases, first relaxation to the steady state of the integrable theory will occur, followed by decay to the thermal ensemble.
Numerical results in \cite{Kollath2007-98} may be interpreted in this vein.

{T.~B.} thanks the DFG and the Studienstiftung des deutschen Volkes for support.

\bibliographystyle{prsty} 

\end{document}